\documentclass[useAMS,usenatbib]{mn2e}
\usepackage{graphics}
\usepackage{epsfig}
\usepackage{amsmath}

\begin{document}

\title{Accreted versus {\it In Situ} Milky Way Globular Clusters}
\author{Duncan A. Forbes and Terry Bridges}
\date{\today}
\maketitle

\begin{abstract}

A large, perhaps dominant fraction, of the Milky Way's halo is
thought to be built-up by the accretion of smaller galaxies and
their globular cluster (GC) systems. Here we examine the Milky Way's
GC system to estimate the fraction of accreted versus {\it in
situ} formed GCs. We first assemble a high quality database of
ages and metallicities for 93 Milky Way GCs from literature deep
colour-magnitude data. The age-metallicity relation for the 
Milky Way's GCs reveals two distinct tracks -- one with near
constant old age of $\sim$12.8 Gyr and the other branches to
younger ages. We find that the latter young track is
dominated by globular clusters associated with
the Sagittarius and Canis Major dwarf galaxies. Despite being
overly simplistic, its
age-metallicity relation can be well
represented by a simple closed box model with continuous star
formation. The inferred chemical enrichment
history is similar to that of the Large Magellanic Cloud, but is more enriched, at
a given age, compared to the Small Magellanic Cloud. 

After excluding Sagittarius and Canis Major GCs, several young
track GCs remain. Their horizontal branch morphologies 
are often red and hence classified as Young Halo objects, however
they do not tend to reveal extended horizontal branches (a
possible signature of an accreted remnant nucleus). Retrograde
orbit GCs (a key signature of accretion) are commonly found in
the young track. We also examine GCs that lie close to the
Fornax-Leo-Sculptor great circle defined by several satellite
galaxies. We find that several GCs are consistent with
the young track and we speculate
that they may have been accreted along with their host dwarf
galaxy, whose nucleus may survive as a GC.
Finally, we suggest that 27-47 GCs (about 1/4 of the entire
system), from 6-8 dwarf
galaxies, were accreted to build the Milky Way GC system we seen today.

\end{abstract}

\section{Introduction}

In a $\Lambda$CDM Universe, Milky Way-like 
galaxy halos are built up by the accretion of
smaller galaxies (Bullock \& Johnson 2005; Abadi et al. 2006;
Font et al. 2006). 
These small galaxies contain dark matter,
stars, gas and in some cases globular clusters (GCs). 
Although the dwarf galaxy may be completely disrupted, the high
density of some GCs allows them to survive the accretion process
(Penarrubia, Walker \& Gilmore 2009). They will then 
add to any pre-existing globular cluster 
population (Searle \& Zinn 1978; Abadi et al. 2006). 
However, galaxy halos
will also include a contribution from stars and clusters 
formed {\it in situ} within the host galaxy. Simulations suggest
that the transition from an {\it in situ} to accretion-dominated
stellar halo  
occurs at a galactocentric radius of $\sim$ 20 kpc for a Milky
Way-like galaxy (Abadi et al. 2006; 
Zolotov et al. 2009). 

In the recent simulation of four Milky Way-like halos, Zolotov et
al (2009) found the accreted fraction of the halo stellar mass to
be 30-85\%, with the large variation due to differences in the 
merger history of the four halos. 
These accreted
stars tended to be acquired early on, and are mostly in place 9 Gyr
ago. The {\it in situ} stars, on the other hand, generally formed at
very early epochs and are more than 11.5 Gyr old. 

Observationally, there is now good evidence for the accretion and
disruption of dwarf galaxies in the form of several 
stellar streams having been detected (Helmi et al. 1999; Newberg et al. 2002;
Duffau et al. 2006; Grillmair 2009), 
the most famous of which is associated with the Sagittarius dwarf
spheriodal galaxy (Ibata, Gilmore \& Irwin 1994). Recently, Carollo et
al. (2007) have presented evidence from the SDSS for a dual
nature of our Galaxy's inner and outer 
halo; with the outer halo indicating more
retrograde motions and lower metallicities indicative of the 
accretion of low-mass galaxies. Streams, from disrupted dwarfs,
have also been detected in the halo of the Andromeda galaxy
(McConnachie et al. 2009).

Perhaps the `gold standard' in searching for evidence of
disrupted dwarf galaxies and their GCs, is via integrals of
motion. These energy and angular momentum parameters 
are largely invariant to phase mixing and hence survive
intact for many orbits (Helmi \& de Zeeuw 2000). However, their
use is limited as the full orbital information is often not 
available and assumptions about the Galactic potential are
required. A phase-space search is a more practical method,
although it is limited by the accuracy of the orbital model of
the disrupted object. Integrations backwards and forwards in
time become increasingly less certain.

Alternative signatures of
accreted GCs may include the Oosterhoff type (Catelan 2005),
peculiar element ratios (Pritzl et al. 2005), ellipticity, size and
luminosity (Mackey \& van den Bergh 2005), and horizontal branch
class and kinematics (Mackey \& Gilmore 2004). 
Despite the strong evidence for
accretion, observational estimates of the mass fractions involved
are difficult to quantify. 
From their study of the Milky
Way GC system, Mackey \&
Gilmore (2004) suggested that 41 (27\%) of the Milky Way's GCs were
accreted and their host dwarf galaxies were responsible for
roughly half of the current halo stellar mass. Using blue halo
stars, Unavane, Wyse \& Gilmore (1996) suggested an accretion fraction
closer to 10\%.

The age-metallicity relation (AMR) of a galaxy can also be used
to constrain its evolutionary history via the  
chemical enrichment of its 
constituent stars and clusters. The exact shape of the AMR is
determined by the complex interplay between Iron injected into
the ISM by SN
Ia and Iron lost/gained via any outflows/accreted gas (e.g
Lanfranchi \& Matteucci 2004). These are
in turn affected by the depth of the potential well and
tidal interactions with another galaxy or the intragroup
medium. For example, interactions can lead to a burst of star
formation as gas is funneled into the galaxy centre (Mayer et
al. 2001; Forbes et al. 2003; Mayer 2009). Tidal interactions
between the Magellanic Clouds and the Milky Way may have been
responsible for the formation of 
young/intermediate age star clusters in the
Clouds (Bekki et al. 2004). 

For `massive' Local Group dwarfs, such as Sgr, the review of Mayer
(2009) concludes they are largely shaped by tidal effects. Over a
few Gyr, tidal shocks induce an instability which drives new star
formation. The outer layers of the galaxy can be lost before
complete disruption.
Previous work on the Sgr dwarf galaxy has shown that it
continued to form field stars and star clusters as it interacted,
and ultimately accreted, with the Milky Way over a period of 
several Gyrs (Layden \& Sarajedini 2000;
Siegel et al. 2007). Models of the Sgr orbit also suggest a long
orbital timescale (Helmi \& White 2001; Fellhauer et al. 2006). 
In Forbes et al. (2004) we showed that the Sgr dwarf
and the accreted Canis Major (CMa) dwarf galaxy revealed a very
similar, well-defined AMR. It is therefore a powerful tool in 
which to identify accreted GCs via their measured 
age and metallicity.

In this work we compile a new catalogue of age and metallicity
measurements for Milky Way GCs from published analysis of deep colour-magnitude
diagrams. We use this catalogue to 
revisit the AMR of the Sgr and CMa dwarfs.
After removing the GCs that can be confidently associated with
these two `accretion events' we proceed to examine the AMR of
the remaining Milky Way GCs. In particular, we focus on those
with retrograde motions, extended horizontal branches and those 
associated with the Fornax-Leo-Sculptor great circle. Finally, we
estimate the accreted vs {\it in situ} fractions for Milky Way
GCs.

\section{A Database of Globular Cluster Ages and Metallicities}

New {\it relative} ages for 64 Milky Way GCs have recently been
determined using deep ACS imaging from the {\it Hubble Space
Telescope} (HST) by Marin-Franch et al. (2009). These data were
sufficiently deep to derive ages from the main sequence turnoff
and form a high quality homogeneous database.
The mean {\it absolute} age of the metal-poor GCs is 12.80 Gyr (using
the Dartmouth 
models of Dotter et al. 2007), which we adopt as the normalisation
to derive absolute ages. Note, that in some cases the relative
age is
greater than the normalisation value, 
with the oldest absolute age in their sample being
14.46 Gyr (albeit with a large error of $\pm$ 1.8 Gyrs). The typical
error quoted is closer to $\pm$ 0.5 Gyr. We use Carretta \& Gratton
(1997; CG) based [Fe/H] metallicities as given by Marin-Franch et al. but
we do not convert these into [M/H] as they have done.

The main limitation of the ACS study above was that the GCs were restricted
to be within $\sim$ 20 kpc of the Galactic centre. We have chosen
to supplement these data with the relative ages of 13 GCs from de
Angeli et al. (2005), again assuming a normalisation age of 12.8
Gyr. Marin-Franch et al. show that their ages are consistent, within
errors, with the full sample of de Angeli et al. Absolute ages and
metallicities for a further 10 GCs are taken from Salaris \& Weiss
(1998). The typical age error here is larger at $\pm$ 1.3 Gyr.

The above data are supplemented by high quality colour-magnitude
diagram studies of individual GCs.  
The age and metallicity for the  
GCs Whiting~1 and AM~4 are taken directly from Carraro et al. (2007) and 
Carraro (2009) respectively. For NGC 5634, 
Bellazzini, Ferraro \& Ibata (2002) 
derived a CG
metallicity of [Fe/H] = --1.94 and an age similar to that of the GCs
NGC 4590 and Terzan 8. Taking the average 
of the Marin-Franch et al. ages for these 
two GCs, i.e. 11.52 and 12.16 Gyr respectively gives 11.84 Gyr
for NGC 5634. We also include two outer GCs from the deep HST
study of AM~1 and Pal~14 from Dotter, Sarajedini \& Yang
(2008). We use their age estimates and convert their
Zinn-West metallicities using Carretta \& Gratton
(1997) to get [Fe/H] = --1.47 for AM~1 and --1.36 for
Pal~14. Catelan et al. (2002) derive a CG metallicity of [Fe/H] =
--1.03 for NGC 6864 (M75) and an age coeval with NGC 1851,
i.e. 9.98 Gyr.     
Thus our final sample of GCs with 
age and metallicity 
measurements totals 93 and is dominated by the recent ACS data.

In Fig. 1 we show the age-metallicity distribution of our sample of 93 GCs, 
coded by 
the different data sources. The Salaris \& Weiss (1998) data are on average 
younger than the other data sources, however this is partly due to the 
fact that the GCs included from their survey 
tend to be `young halo' objects at relatively 
large galactocentric distances. The distribution shows a relatively constant 
old age for the most metal-poor GCs. For the ACS data,
Marin-Franch et al. (2009) measured a rms dispersion of only $\pm$0.64
Gyr for the most metal-poor GCs. 
At [Fe/H] $\sim$ --1.5 there is a track 
to younger ages (which is particularly tight if only the ACS data
are considered) and, after a small gap in metallicity, 
a group of metal-rich GCs at old ages. The latter group 
are largely associated with the bulge of the Milky Way 
(see Minniti 1995 and C\^{o}t\'{e} 1999 for further details), although
some may be inner halo objects (Burkert \& Smith 1997). 

Our final sample includes almost 2/3 of the known GCs in the
Milky Way. In Fig. 2 we compare the metallicity distribution for
148 GCs from the latest version of the Harris (1996)
catalogue with our final sample. 
Although we have converted the Harris quoted metallicities from the 
Zinn \& West scale to the Carretta \& Gratton (1997) scale,
slight differences will exist for individual GCs. Nevertheless,
the histograms show that we sample well the low metallicity end
of the distribution but we are under-represented by metal-rich
GCs. A number of these GCs are located in the bulge, and the lack
of good age determinations is probably due to the difficulty of obtaining deep
colour-magnitude diagrams (due to foreground stars and
extinction).

\section{Sagittarius and Canis Major Dwarf Galaxies}

\subsection{Sagittarius (Sgr) dwarf galaxy}

Since its initial discovery by Ibata, Gilmore \& Irwin 
(1994), several GCs have been associated with the Sgr dwarf galaxy on the 
basis of their location in phase space compared to the predicted orbit of 
the dwarf (Bellazzini et al. 2003). They are 
Terzan 7, Terzan 8, Arp 2, Pal 12, NGC 4147 
and NGC 6715 (M54), with the latter being the 
probable remnant nucleus (see also Bellazzini et al. 2008). To this list 
of 6 GCs we include the GC Whiting~1, which Carraro, Zinn \& Moni Bidin 
(2007) have shown is consistent with the Sgr orbit in phase space. 

Carraro \& Bensby (2009) recently concluded that the outer disk open 
clusters (OCs), 
Berkeley~29 and Saurer~1, were also 
associated with the Sgr dwarf in phase space.
Here we use the age and metallicity of these open clusters from 
their table 1. 
The Sgr dwarf galaxy reveals ongoing star formation in its field star 
population (Layden \& Sarajedini 2000). Siegel et al. (2007)
have derived 
new age and metallicity estimates for the Sgr field stars using the HST/ACS.
We include their young/intermediate age field star populations. 

Thus in revisiting the age-metallicity relation for the Sgr dwarf 
galaxy, we have updated data for 7 GCs, 2 OCs and 3 field star populations.

\subsection{Canis Major (CMa) dwarf galaxy} 

Although its existence as a disrupted dwarf galaxy and its connection
to the Monoceros overdensity is still the subject of debate (Mateu et
al.  2009), there is good evidence supporting its distinct nature
(Martin et al. 2004; Dinescu et al. 2005). 
Martin et al. (2004) found that four GCs (NGC
1851, 1902, 2298, 2808) occupied a tight location in phase space,
close to their orbital model for CMa.  
Bellazzini et al. (2004) argued
on the basis of position, distance and stellar populations that the
old open clusters AM~2 and Tom~2 were also associated with the CMa
dwarf. This was supported by F04.  The field stellar
population of CMa is currently less well constrained than for the Sgr
dwarf, so we do not consider field stars further.  
We consider 4 GCs and 2 OCs initially associated with the CMa
dwarf.

\subsection{A combined Sgr and CMa AMR}

In F04 we showed that the AMR for the Sgr and CMa dwarfs, as defined by
their GCs, OCs and field stars, were very similar. In Fig. 3
we plot the open and globular clusters that can be confidently 
associated with these two disrupted dwarf galaxies using updated
data as described above. We also include the data points for the
young/intermediate field star populations in the Sgr dwarf from the
HST/ACS study of Siegel et al. (2007). Finally, we include NGC
5634 and AM~4 as probable members of the Sgr dwarf, and NGC 4590,
Pal~1 and Rup~106 as probable members of the CMa dwarf (see
Section 3.4 and Table 1). 


Fig. 3 shows that the updated data for the Sgr and CMa dwarfs continue
to reveal a similar AMR for both galaxies which is remarkably tight 
over a large range in
age and metallicity. This further supports claims that CMa is indeed a
distinct feature (i.e. a remnant dwarf galaxy) and not simply an
overabundance or warp in the Galactic disk.  Both galaxies appear to
have been forming stars and star clusters for $\sim$10 Gyr enriching from 
about 1/100 solar to solar metallicity.

We also show a simple closed box chemical enrichment model, with constant 
star formation from a starting age of 12.8 Gyr. A 
single AMR provides a good representation for the chemical enrichment 
history of both dwarf galaxies.  
We show a dispersion about the AMR of $\pm$ 0.64 Gyr which is the
same as the rms dispersion of the metal-poor GCs in the ACS study
of Marin-Franch et al. (2009). 
We note
that the Sgr and CMa dwarfs have a similar luminosity of M$_V$ $\sim$
--14 (Bellazzini et al. 2006). So as well as a similar mass, they appear 
to have had a similar enrichment history. Below we refer to the combined 
Sgr and CMa AMR as the joint AMR. 

The AMR of the Sgr dwarf has been modelled by Lanfranchi \&
Matteucci (2004), along with other Milky Way dwarf
satellites. They suggested that Sgr had a particularly high star
formation efficiency compared to other dwarfs
and its chemical enrichment has proceeded
continuously since its formation some 13 Gyr ago. Our results,
combining field stars and clusters, supports this model.

\subsection{Additional Sgr and CMa GCs}

Given the tightness of the joint AMR we can use it as 
a diagnostic for helping to confirm the 
membership of other GC candidates that have been suggested on the 
basis of their location in phase space.

Carraro (2009) has suggested a possible association of the GC AM~4 with the 
Sgr dwarf although it lies at a somewhat higher galactic latitude than the 
Law et al. (2005) orbital model.  
As shown in Fig. 4, the age and metallicity of AM~4 are quite distinct from 
the general GC distribution but they 
are consistent with the 
joint AMR. Thus we support the suggestion of Carraro (2009) that AM~4 is 
associated with the Sgr dwarf. 

Bellazzini, Ferraro \& Ibata (2002) investigated the phase space
location of NGC 5634 suggesting it may
belong to the Sgr GC system.  
It is a metal-poor GC but 
somewhat younger than the general GC distribution and we tentatively 
support Bellazzini et al. claim and include NGC 5634 in the Sgr GC 
system.

The faint (M$_V$ $\sim$ --2) GC Koposov~1 is spatially nearby to 
stars from the Sgr dwarf which led Koposov et al.~(2007) 
to suggest an association. They quote stellar population 
estimates of $\sim$8 Gyrs and [Fe/H] $\sim$ --2. Such values deviate strongly 
from the joint AMR and so we conclude that Koposov~1 is very unlikely 
to be a member of the Sgr dwarf GC system. 

Martin et al. (2004) 
identified several GCs that lie close to the favoured prograde orbit of 
the CMa dwarf in 
phase space (but see also Penarrubia et al. 2005 for alternative
view). 
These potential CMa GCs are NGC 4590, 5286, 6205, 6341, 6426,
6779, 7078, 7089, IC 1257, Pal~1 and Rup~106. Our sample includes all 
except IC 1257. We show the remaining 10 GCs in Fig. 4, along with 
the joint AMR and main GC distribution. 
Of these GCs we support the association of NGC 4590, Pal~1 and
Rup~106 
based on their
consistency with the joint AMR. We note that Pritzl et
al. (2005) favours an extragalactic origin for NGC 4590 and
Rup~106 on the basis of their peculiar
alpha-element content. 
For the other 7 GCs, they 
may have an association with CMa 
but this can not be determined on the basis of their 
stellar populations as they 
overlap with both the joint AMR and the general AMR. So 
for now, we only include NGC 4590, Pal~1 and Rup~106 with the CMa dwarf. 

Thus, we have added AM~4 and NGC 5634 to the Sgr dwarf GC system and 
NGC 4590, Pal~1 and Rup~106 to the CMa dwarf GC system. These
GCs were included in the joint AMR shown in Fig. 3. 

As summarised in Table 1, we assign 9 GCs to the Sgr dwarf and 7 to
CMa dwarf. The specific frequency S$_N$ assuming M$_V$ = --14 for both
galaxies would be 23 and 18 respectively. If we associate one GC with
the nucleus of each dwarf (NGC 6715 for Sgr and NGC 2808 for CMa), then the
S$_N$ values would become 20 and 15. Although such values are high, they  
are similar to that for the Fornax dSph galaxy which has a system of 5 GCs 
for a luminosity of M$_V$ $\sim$ --13, giving S$_N$ $\sim$ 30. Other 
dwarf galaxies also reveal high S$_N$ values (e.g. Durrell et al. 1996). 
We also note that high frequency values for dwarfs are less extreme
when one considers the total GC mass normalised by the galaxy
stellar mass (e.g. see Spitler \& Forbes 2009). 


Tables 2 and 3 list 
the age and metallicity of the remaining Milky Way GCs in our 
sample that we have not associated with either the Sgr or CMa dwarf 
galaxies. Most of the `young track' GCs are no longer present,
however there are still several GCs at intermediate metallicities
with younger than average ages. These are prime candidates for
accreted GCs, which we discuss further below. 

\begin{table}
\caption{Sagittarius and Canis Major Star Clusters.}
\begin{tabular}{lccrcl}
\hline
Name & Age & [Fe/H] & Source & Class & Member\\
& (Gyr) & (dex) & & & \\
\hline
{\it Sagittarius}  & & & &\\
NGC~6715 &      10.75 & --1.25 & 1 & YH & $\surd$ $\surd$\\
Terzan~7 & 7.3 & --0.56 & 1 & YH & $\surd$ $\surd$\\
Terzan~8 & 12.16 & --1.80 & 1& OH & $\surd$ $\surd$\\
Arp~2 &    10.88 & --1.45 & 1 & YH & $\surd$ $\surd$\\
Pal~12 &    8.83 & --0.83 & 1 & YH & $\surd$ $\surd$\\
NGC~4147 & 11.39 & --1.50 & 1 & YH & $\surd$ $\surd$\\
Berk~29 & 4.5 & --0.44 & 7 & OC & $\surd$ $\surd$\\
Saurer~1 & 5.0 & --0.38 & 7 & OC & $\surd$ $\surd$\\
Whiting~1 & 6.5 & --0.65 & 4 & UN &  $\surd$ $\surd$\\
AM~4 & 9.0 & --0.97 & 5 & UN & $\surd$\\
NGC~5634 & 11.84 & --1.94  & 6 & OH & $\surd$ \\
\hline
{\it Canis Major} & & & & \\
NGC~1851 & 9.98 & --1.03 & 1 & OH & $\surd$ $\surd$\\
NGC~1904 & 11.14 & --1.37  & 2 & OH & $\surd$ $\surd$\\
NGC~2298 & 12.67 & --1.71 & 1 & OH & $\surd$ $\surd$\\
NGC~2808 & 10.80 & --1.11 & 1 & OH & $\surd$ $\surd$\\
AM~2 & 5.0 & --0.50  & 8 & OC & $\surd$ $\surd$\\
Tom~2 & 2.0 & --0.28  & 9 & OC & $\surd$ $\surd$\\
NGC~4590 & 11.52 & --2.00 & 1 & YH & $\surd$\\
Pal~1 & 7.30 & --0.70 & 1 & BD & $\surd$\\
Rup~106 & 10.2 & --1.49 & 3 & YH & $\surd$\\
\hline
\end{tabular}
\\
Notes: YH = young halo, OH = old halo, BD = bulge/disk, UN = unknown, OC = 
open cluster. $\surd$ $\surd$ = secure member on the basis of phase space, 
$\surd$ = probable member. 
1 = Marin-Franch et al. (2009); 2 = de Angeli et al. (2005); 3 = 
Salaris \& Weiss (1998); 4 = Carraro et al. (2007); 5 = Carraro (2009); 
6 = Bellazzini et al. (2002); 7 = Carraro \& Bensby (2009); 8 = 
Lee (1997); 9 = Frinchaboy et al. (2008).  
\end{table}

\section{Magellanic Clouds}

The small and large Magellanic (SMC, LMC) clouds are currently interacting
with the Milky Way. This interaction may have served to dislodge
some Magellanic cloud GCs, which we now associate with the Milky
Way. 

In Fig. 5 we compare the age-metallicity relations of the SMC
and LMC with that of the joint AMR and the general
distribution of Milky Way GCs. We include 
the average of the disk and bar region in the LMC 
from Carrera et al. (2008a) and
the SMC from Carrera et al. (2008b). Metallicities come from
Calcium triplet measurements on the Carretta \& Graton (1997)
scale. We also include the 7 SMC open and globular clusters from the
ACS study by Glatt et al. (2008). We take their derived ages using
Dartmouth isochrones (Dotter et al. 2007) and we convert their
metallicities to the Carretta \& Graton (1997) scale. We also
include the `age-gap' GC in the LMC from ACS data 
by Mackey, Payne \& Gilmore (2006), taking their age estimate of
73\% of the age for NGC 104 (47 Tuc).  
Both the SMC and LMC AMRs can be
well represented by a simple closed-box model (Carrera et
al. 2008a,b; Glatt et al. 2008). 

The figure shows that the LMC has a similar AMR to that of 
Sgr and CMa dwarfs.   
The SMC AMR is systematically shifted to younger ages for a
given metallicity. 
We conclude on this basis that it is very unlikely that
Rup~106 (age = 10.2 Gyr, [Fe/H] = --1.49) 
was formerly an LMC member as suggested by Lin \& Richer
(1992). However, it may be associated with the SMC as suggested Fusi Pecci et
al. (1995), or with the CMa dwarf as discussed above.

\section{Zinn Horizontal Branch Morphology}

Zinn (1993) separated the halo GCs on the basis of their
horizontal branch (HB) morphology at a 
given metallicity, and assumed that this effect was driven by
age. 
The so-called `young halo' GCs have red HBs at fixed 
metallicity, tend to be located at large galactocentric radii and reveal 
a mean retograde motion in contrast to the prograde orbits of the
`old halo'. These properties have led many to suggest 
that the young halo GCs have been accreted (e.g. van den Bergh
1993). On the basis of their 
spatial distribution and sizes, Mackey \& Gilmore
(2004) suggested that $\sim$16\%, or $\sim$11, of the old halo GCs were 
also accreted. 

Using the classification 
of either bulge/disk (BD), young halo (YH) or 
old halo (OH) as listed by Mackey \& van den Bergh (2005), we show the
age-metallicity diagram coded by HB class in Fig. 6. We also
include 
NGC 5139 (Omega Cen) that Catelan (2005) list a HB ratio of +0.89
for, which indicates an 
OH classification for its metallicity. 

The figure shows a clear separation between BD GCs at high metallicity 
and the halo GCs. The young halo GCs indeed have younger ages on average  
compared to old halo GCs.  
A couple of GCs classified as OH are located at younger ages than typical OH 
GCs (i.e. NGC 6712 and NGC 6535). 

If we assume that all of the OH GCs that follow the joint AMR are
accreted, then the number is about a dozen GCs, consistent with
the estimate of Mackey \& Gilmore (2004). 

Mackey \& Gimore (2004) found that YH GCs shared many traits of
GCs in nearby dwarf galaxies, however 
we find that the Sgr and CMa GCs reveal a range of HB classes (see Table 1). 
The Sgr dwarf contains 
mostly YH GCs but also includes two OH GCs (Terzan~8 and NGC 5634). The CMa 
dwarf GCs are mostly OH, except NGC 4590 (YH) and Rup~106 
(YH) and Pal~1 (BD). Thus the 
GCs which we are most 
confident were accreted, cover a range of HB classes. 
This suggests HB class is only a crude accretion indicator at best.

\section{Extended Horizontal Branch (EHB) Globular Clusters}

EHBs in GCs appear to be due to the presence of a helium-enhanced
population of stars and hence are a strong indicator for the presence of
multiple stellar populations (Norris 2004; Piotto et al. 2005). Such
multiple populations could be the result of enhanced enrichment in the
nuclei of dwarf galaxies which have subsequently been stripped and
accreted (Lee et al. 2007; Bekki et al. 2007). Lee et al. (2007)
determined that 15 out of 114 (13\%) 
GCs examined have evidence for a strong EHB. 
They are: NGC 2419, 2808, 5139 (Omega Cen), 5986, 6093, 6205, 6266, 6273, 
6388, 6441, 6656, 6715 (M54), 6752, 7078 and 7089. NGC 2419 and
5139 (Omega Cen) have previously been associated with dwarf
galaxy nuclei (Mackey \& van den Bergh 2005).  

After excluding the Sgr `nucleus' (NGC 6715) and the suspected 
equivalent for CMa (NGC 2808), the age and metallicity of the remaining 
13 EHB GCs are shown in Fig. 7. The figure shows that the EHB
GCs have a wide range of metallicities. They are on 
average slightly younger with ages 11-12 Gyr, for a given metallicity, than the
general distribution. This may be a reflection of age as the second
parameter effect.
Otherwise they are consistent with
both the 
joint and general AMRs. Without the presence of associated young
GCs, it is difficult to identify possible accreted GCs in a sample of
EHBs using an age-metallicity analysis.

\section{Retrograde Motion GCs}

GCs on retrograde orbits are prime candidates for having been accreted 
from a satellite on an initially retrograde encounter with the 
Milky Way. Perhaps the most well known retrograde motion case is NGC 5139 
(Omega Cen). 
Taking into account the potential of the Milky Way bar, Allen et al. 
(2006, 2008) calculated the z component of the orbital angular momentum 
for 54 GCs (about 1/3 of the Milky Way GC system). They found 16/54
(30\%) to be on retrograde orbits for their model 
of the Milky Way potential. They are: NGC 288, 362, 2298, 3201,
5139 (Omega Cen), 5466, 6121, 6144, 6205, 6316, 6712, 6779, 6934, 7089, 
7099 and Pal~13.
All are available in our sample except NGC 6316 and Pal~13. 
We note that none of these 16 GCs belong to the CMa or Sgr systems, except 
NGC 2298. However the detection of retrograde motion for NGC 2298 is 
of low significance and we continue to include it in the CMa GC sample. 
The location of the remaining 13 GCs with respect to the 
joint AMR are shown in Fig. 8. Note that for many of the other GCs shown in 
this plot the orbital properties are unknown and remain unclassified.

Five of the retrograde GCs are consistent with both the joint and
general AMRs. However several 
are only well-described by the joint AMR. They are 
NGC 288, 362, 3201, 5139 (Omega Cen) 6205, 6712, 6934, 7089.

NGC 5139 (Omega Cen) is thought to be the remnant nucleus of a
dwarf galaxy (Freeman 1993). Does it have associated GCs? 
Dinescu et al. (1999) noted that NGC 362 and NGC 6779 shared a similarly
strong retrograde motion with Omega Cen for a low inclination. Remarkably both 
Omega Cen and NGC 362 lie directly on the joint AMR, whereas NGC
6779 is very metal-poor. According to the 
best fit models of Bekki \& Freeman (2003), the progenitor dwarf galaxy 
had a luminosity of M$_V$ $\sim$ --14.5. This is similar to the predicted 
luminosity of both the Sgr and CMa dwarf galaxies. Although,
galaxy mass is only one of many parameters that may determine a galaxy's AMR
(Lanfranchi \& Matteucci 2004), we tentatively support the
suggestion that NGC
362 and NGC 6779 may be associated with the disrupted dwarf
galaxy whose nucleus is now NGC 5139 (Omega Cen). We conclude
that a retrograde orbit is a useful discriminant of an accreted
GC and that larger samples of GCs with such information will be a
useful addition to Galactic archaeology studies.

\section{Fornax-Leo-Sculptor Great Circle}

A great circle in the sky connects the Fornax, Leo (I and II)
and Sculptor (FLS) galaxies (Lynden-Bell \& Lynden-Bell 1995). Several
recently discovered dwarf galaxies lie within a few degrees of
the FLS great circle. They include Coma, CVn II, CVn 
and the disrupting GC Segue I. The dwarf Sextans also lies nearby.
The FLS great circle given by Fusi Pecci et al. (1995) passes close to several
GCs. Once the GCs associated with CMa dwarf are excluded, the 11
GCs nearest to the orbit are: NGC 1261, 6426, 6864, Pal~3,
Eridanus, Pal~14, NGC 5053, 5024, AM~1, Pal~15 and Pal~4. Our sample
includes all except Pal~15. 

In Fig. 9 we show the location of these 10 GCs, along with the
$\sim$10 Gyr old field stars in the Fornax galaxy from Coleman \&
de Jong (2008). Seven of the 10 GCs reveal ages younger by $\sim$3 Gyr
than the general distribution, and consistent with the field
stars that formed $\sim$10 Gyr ago in the Fornax galaxy. These GCs
account for most of the remaining intermediate metallicity
objects that deviate from the general distribution. 

One possibility is that some of these GCs were tidally stripped from
the Fornax galaxy as it crossed the orbit of the LMC (Dinescu et
al. 2004). 
A second possibility is that these GCs came from the remains of a
completely disrupted dwarf galaxy that shared the stream with
half a dozen other satellites. The remnant nucleus of this
possible dwarf may be NGC 5024 or NGC 5053, which are  
the 4th and 6th largest GCs for their luminosity
respectively. The largest three 
are NGC 5139 (Omega Cen), NGC 6715 (M54) and NGC
2419 which are all potential remnant nuclei of dwarf galaxies
(Mackey \& van den Bergh 2005). We note that NGC 5053 reveals 
evidence of a tidal stream (Lauchner et al. 2006). 
Unfortunately, most of the potential FLS GCs do not have published
proper motions and so their integrals of
motion are not available. Exploring these GCs further requires
more data.


\begin{table}
\caption{Milky Way Globular Clusters.}
\begin{tabular}{lrrc}
\hline
Name & Age & [Fe/H] & Source\\
& (Gyr) & (dex) & \\
\hline
     NGC~0104  &     13.06   &     -0.78 & 1\\
     NGC~0288  &     10.62   &     -1.14 & 1\\
     NGC~0362  &     10.37   &     -1.09 & 1\\
     NGC~1261  &     10.24   &     -1.08 & 1\\
     NGC~2419  &      12.3   &     -2.14 & 3\\
     NGC~3201  &     10.24   &     -1.24 & 1\\
     NGC~4372  &     12.54   &     -1.88 & 2\\
     NGC~4833  &     12.54   &     -1.71 & 1\\
     NGC~5024  &     12.67   &     -1.86 & 1\\
     NGC~5053  &     12.29   &     -1.98 & 1\\
     NGC~5139  &     11.52   &     -1.35 & 1\\
     NGC~5272  &     11.39   &     -1.34 & 1\\
     NGC~5286  &     12.54   &     -1.41 & 1\\
     NGC~5466  &     13.57   &     -2.20 & 1\\
     NGC~5694  &     13.44   &     -1.74 & 2\\
     NGC~5824  &     12.80   &     -1.60 & 2\\
     NGC~5897  &      12.3   &     -1.73 & 3\\
     NGC~5904  &     10.62   &     -1.12 & 1\\
     NGC~5927  &     12.67   &     -0.64 & 1\\
     NGC~5946  &     11.39   &     -1.22 & 2\\
     NGC~5986  &     12.16   &     -1.35 & 1\\
     NGC~6093  &     12.54   &     -1.47 & 1\\
     NGC~6101  &     12.54   &     -1.76 & 1\\
     NGC~6121  &     12.54   &     -1.05 & 1\\
     NGC~6144  &     13.82   &     -1.56 & 1\\
     NGC~6171  &     13.95   &     -0.95 & 1\\
     NGC~6205  &     11.65   &     -1.33 & 1\\
     NGC~6218  &     12.67   &     -1.14 & 1\\
     NGC~6235  &     11.39   &     -1.18 & 2\\
     NGC~6254  &     11.39   &     -1.25 & 1\\
     NGC~6266  &     11.78   &     -1.02 & 2\\
     NGC~6273  &     11.90   &     -1.53 & 2\\
     NGC~6284  &     11.14   &     -1.13 & 2\\
     NGC~6287  &     13.57   &     -1.91 & 2\\
     NGC~6304  &     13.57   &     -0.66 & 1\\
     NGC~6341  &     13.18   &     -2.16 & 1\\
     NGC~6342  &     12.03   &     -0.69 & 2\\
     NGC~6352  &     12.67   &     -0.70 & 1\\
     NGC~6362  &     13.57   &     -0.99 & 1\\
     NGC~6366  &     13.31   &     -0.73 & 1\\
     NGC~6388  &     12.03   &     -0.77 & 1\\
     NGC~6397  &     12.67   &     -1.76 & 1\\
     NGC~6426  &      12.9   &     -2.11 & 3\\
     NGC~6441  &     11.26   &     -0.60 & 1\\
     NGC~6496  &     12.42   &     -0.70 & 1\\
     NGC~6535  &     10.50   &     -1.51 & 1\\
     NGC~6541  &     12.93   &     -1.53 & 1\\
     NGC~6544  &     10.37   &     -1.20 & 2\\
     NGC~6584  &     11.26   &     -1.30 & 1\\
     NGC~6624  &     12.54   &     -0.70 & 1\\
     NGC~6637  &     13.06   &     -0.78 & 1\\
     NGC~6652  &     12.93   &     -0.97 & 1\\
     NGC~6656  &     12.67   &     -1.49 & 1\\
     NGC~6681  &     12.80   &     -1.35 & 1\\
     NGC~6712  &      10.4   &     -0.94 & 3\\
     NGC~6717  &     13.18   &     -1.09 & 1\\
     NGC~6723  &     13.06   &     -0.96 & 1\\
     NGC~6752  &     11.78   &     -1.24 & 1\\
     NGC~6779  &     13.70   &     -2.00 & 1\\
\hline
\end{tabular}
\\
Notes: 1 = Marin-Franch et al. (2009); 2 = de Angeli et al. (2005); 3 = 
Salaris \& Weiss (1998); 4 = Dotter et al. (2008); 5 = Catelan et
al. (2002).
\end{table}

\begin{table}
\caption{Table 2 (cont.)}
\begin{tabular}{lrrc}
\hline
Name & Age & [Fe/H] & Source\\
& (Gyr) & (dex) & \\
\hline
     NGC~6809  &     12.29   &     -1.54 & 1\\
     NGC~6838  &     13.70   &     -0.73 & 1\\
     NGC~6864  &     9.98    &     -1.03 & 5\\
     NGC~6934  &     11.14   &     -1.32 & 1\\
     NGC~6981  &     10.88   &     -1.21 & 1\\
     NGC~7078  &     12.93   &     -2.02 & 1\\
     NGC~7089  &     11.78   &     -1.31 & 1\\
     NGC~7099  &     12.93   &     -1.92 & 1\\
     NGC~7492  &      12.0   &     -1.41 & 3\\
        Pal~3 &        9.7  &      -1.39&  3\\
        Pal~4 &        9.5  &      -1.07&  3\\
        Pal~5 &        9.8  &      -1.24&  3\\
        Pal~14 &       10.5 &      -1.36&  4\\
        AM~1  &        11.1 &      -1.47&  4\\
          E3  &     12.80   &     -0.83 & 1\\
    Eridanus  &       8.9   &     -1.20 & 3\\
      Lyng\r{a}~7 &      14.46  &      -0.64&  1\\
\hline
\end{tabular}
\\
Notes: 1 = Marin-Franch et al. (2009); 2 = de Angeli et al. (2005); 3 = 
Salaris \& Weiss (1998); 4 = Dotter et al. (2008); 5 = Catelan et
al. (2002).
\end{table}

\section{Globular Clusters associated with other Streams}


Recently, Newberg et al. (2009) found a thin stream of metal-poor stars 
with a low velocity dispersion that they suggest may have come from 
a progenitor dwarf galaxy. The GC  
NGC 5824 lies nearby in phase space and they suggest it 
may represent the remnant nucleus of the dwarf. Its half light size 
and luminosity place it close to the boundary line of Mackey \& van den 
Bergh (2005). The location of NGC 5824 in Fig. 4 shows its age
(=12.80 Gyr) and 
metallicity (=--1.60) 
are consistent with both the joint AMR and the general GC 
distribution. 

Willett et al. (2009) have searched for GCs associated with their
orbital fit to the Grillmair Dionatos stream, but none were
found.


\section{The number of Accreted Globular Clusters and Dwarf Galaxies}

Previous work has attempted to quantify the number of accreted
dwarf galaxies on the basis of the number of accreted GCs. For
example, 
van den Bergh (2000) suggested 3-7 accreted Sgr-type dwarfs with
4-5 GCs each, for a grand total of $\le$35 accreted GCs. 
Whereas Mackey \& Gilmore (2004) suggested a total of $\sim$41
GCs from $\sim$7 dwarf galaxies. 

From our work, we associate 9 GCs with Sgr dwarf, 7 with CMa
(although some are best described as `probable' 
members), 4-8 GCs with retrograde
motions which also have young ages for their metallicity (this rises to
16 if only their kinematics are considered), and potentially 7
associated with the NGC 5053 disrupted dwarf galaxy. This would
still leave about 8 GCs with young ages for their (intermediate)
metallicities. Thus we suggest 27-47 accreted GCs from 6-8
dwarf galaxies (which suggests that about half of the halo stellar
mass has been accreted in this way). These numbers are 
likely a lower limit, as our sample (and the subsamples discussed
here) represent only a fraction of the known GC system of the Milky Way. 

\section{Conclusions}

We have supplemented new data from
the HST/ACS by Marin-Franch et al. (2009) 
with deep colour-magnitude diagrams from the
literature to
form a high quality database of 93 Milky Way globular clusters
(GCs) with age and metallicity estimates. Additional parameters,
such as magnitude, half-light size, galactoccentric distance 
and horizontal branch class are available for most of the GCs and 
can be found in either Harris (1996) or Mackey \& van den Bergh
(2005). The Milky Way GC age-metallicity relation (AMR) reveals two
distinct tracks -- one has a near constant old age of $\sim$ 12.8
Gyr over the full metallicity range, the other is a track to
younger ages at a given metallicity which appears at intermediate
metallicities. 

Using these new data, we revisit the age-metallicity relations of
two dwarf galaxies thought to have been disrupted and accreted by
the Milky Way -- the Sagittarius (Sgr) and Canis Major (CMa) dwarf galaxies. We
find that the GCs and open clusters associated with the suspected
orbit of each dwarf galaxy produce a well-defined AMR. 
This
supports claims that CMa is a disrupted dwarf and not merely an
overdensity or warp in the Galactic disk.
The joint
Sgr plus CMa AMR indicates that both galaxies produced star
clusters over $\sim$10 Gyr in a manner that 
is consistent with a simple closed box model for
chemical enrichment.
As well as confirming the
secure clusters, we tentatively identify GCs AM4 and NGC 5634 with Sgr, plus
NGC 4590, Pal~1 and Rup~106 with CMa dwarf.  
We find that Koposov~1 is unlikely to a member of the Sgr dwarf
on the basis of its published age and metallicity. For other potential CMa
GCs their age and metallicity did not provide a clear discriminator.
We associated a total of 9 GCs with Sgr and 7 with CMa. The GCs
NGC 6715 (M54) and NGC 2808 (CMa) may be the 
remnant nuclei of the original host galaxy. The GC specific
frequency of each galaxy appears to be quite high at
$\sim$20, but is similar to the Fornax dSph galaxy. 

Comparison with the AMRs of the Large and Small Magllenic Clouds
reveals that the LMC has a similar AMR to the joint Sgr plus CMa
AMR, whereas the SMC is less enriched for a given age.

Once the GCs associated with the Sgr and CMa dwarfs are removed
from the distribution, the remaining GCs are mostly very
old. The remaining sample is dominated by  
GCs formed in a rapid enrichment process on a timescale of less
than 1 Gyr, suggesting that they are largely formed {\it in
situ}.  
However, there are still some GCs with young
ages for their (intermediate) metallicities. These are prime
candidates for having been accreted along with their host dwarf galaxy.

We examined the horizontal branch morphology of GCs as
originally studied by Zinn (1993). Using the recent classification
of Mackey \& van den Bergh (2005) we add NGC 6715 (M54)
with an old halo classification and suggest NGC 6366 should be
modified to bulge/disk. Although old halo GCs are often thought
to be an {\it in situ} subpopulation, we find that several such
GCs have young ages for their metallicity and note that several
of the Sgr and CMa GCs have old halo classifications. Thus
accreted GCs may have old (and young) halo horizontal branch
morphologies. 

Globular clusters with extended horizontal branches have been suggested to be
the remnant nuclei of disrupted dwarfs. After removing NGC 6715
(the probable Sgr nucleus) and NGC 2808 (the possible CMa
nucleus), we found these GCs to be slightly younger on average
but otherwise they covered the full range in metallicity. 

Retrograde motion is the signature of an object that has
been accreted in the opposite rotational sense to the main bulk
of the Galaxy's rotation, and there are 16 known GCs with such
orbits. We find several of these also have young ages for their
metallicity, i.e NGC 288, 362, 3201, 5139, 6205, 6712, 6934 and 
7089. NGC 5139 (Omega Cen) has long been identified as a remnant
nucleus of a dwarf galaxy. Interestingly, it and 
NGC 362 are consistent with an AMR similar to that of Sgr and
CMa. 

It has been known for some time that dwarf galaxies are not
isotropically distributed but rather form 
great circles in the sky. The Fornax-Leo-Sculptor great circle
(which includes several recently discovered faint dwarf galaxies)
passes nearby to several GCs. After excluding the Sgr and CMa
GCs, we examined the
age-metallicity distribution for the 10 GCs nearest to the great
circle. We find 7 of
them to have young ages 
suggesting they were accreted.

On the basis of our age-metallicity analysis, we suggest that at least
27-47 GCs, from 6-8 dwarf galaxies, were
accreted by the Milky Way over the course of its evolutionary
history. 

Currently scheduled for a 2012 launch, the GAIA spacecraft will
determine accurate positions and space motions for most globular
clusters, nearby dwarf galaxies and relic streams of stars. Once
such data are available, we expect it to provide a new impetus to the field of
Galactic archaeology and allow many of the suggestions made here to
be tested.


\section{Acknowledgements}

We thank L. Spitler, C. Foster, D. Hanes, S. van den Bergh, W. Harris 
and J. Strader for useful discussions. DF thanks the ARC for
financial support and Queen's University for hosting his visit.

\newpage

\begin{figure*}
\begin{center}
\includegraphics[scale=0.5,angle=-90]{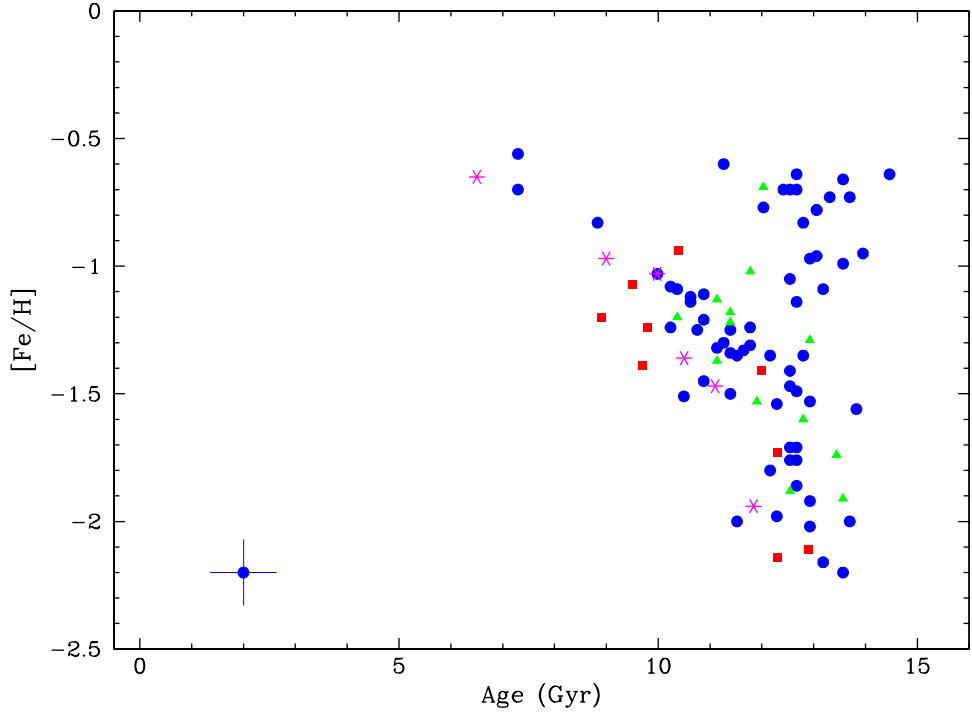}
\caption{Globular cluster sample coded by source of data. 
Marin-Franch et al. (2009) are blue circles, Salaris \& 
Weiss (1998) are red squares, 
de Angeli et al. (2005) are green triangles, Carraro (2009),  
Carraro et al. (2007), Bellazzini et al. (2002), Catelan et
al. (2002) 
and Dotter et al. (2008) are purple stars. A typical error bar for the 
Marin-Franch et al. ACS data is shown lower left. 
}
\end{center}
\end{figure*}

\begin{figure*}
\begin{center}
\includegraphics[scale=0.5,angle=-90]{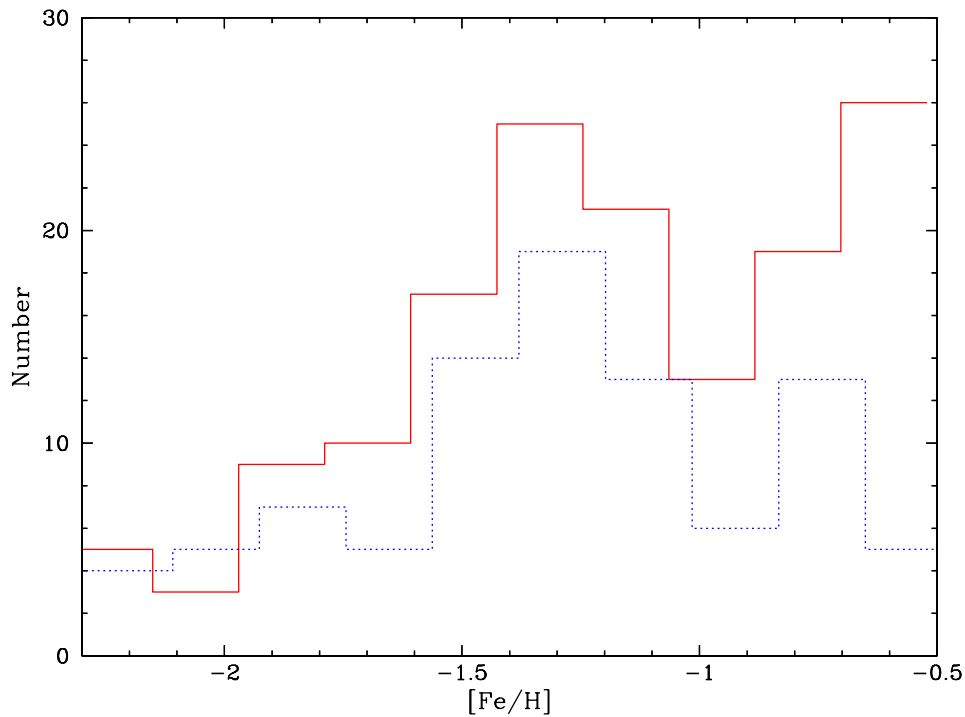}
\caption{Histogram of globular cluster metallicities. The Milky
Way distribution, converted into Carretta \& Gratton (1997)
metallicities, is shown by the solid red histogram. The
distribution of our sample of 93 GCs is shown as a dotted blue
histogram. Our sample is under-represented in metal-rich
GCs compared to the full Milky Way distribution.
}
\end{center}
\end{figure*}

\begin{figure*}
\begin{center}
\includegraphics[scale=0.5,angle=-90]{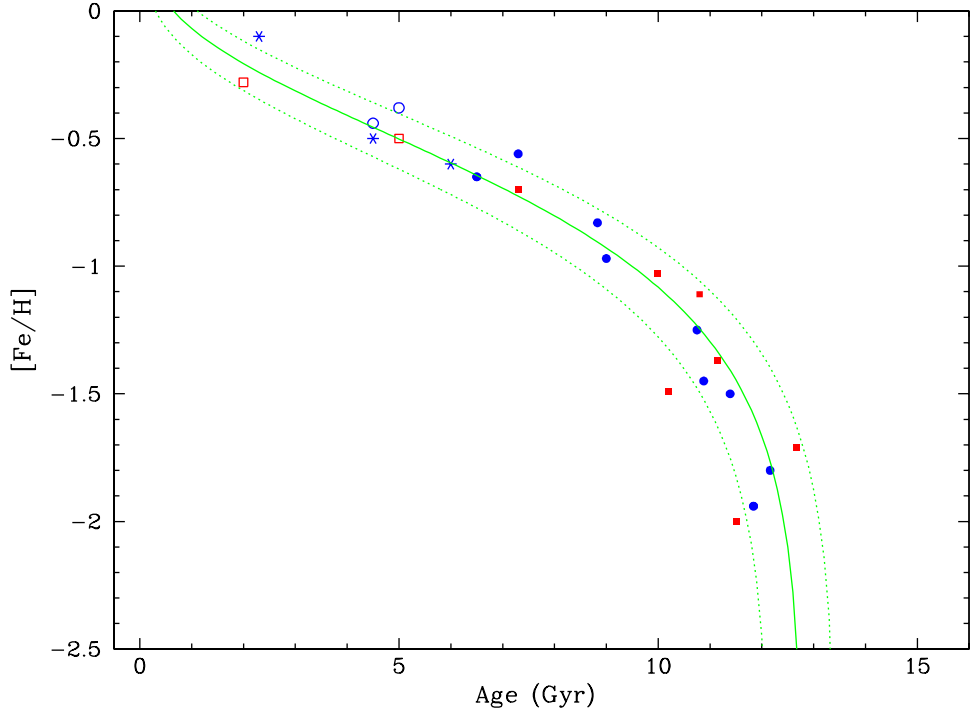}
\caption{Joint age-metallicity relation for the Sgr and CMa dwarf galaxies. 
The Sgr dwarf data are blue circles and the CMa are red squares, 
with open symbols for open clusters  
and stars for field star populations. The green line is a simple closed box 
model with a continuous star formation rate to represent the
joint Sgr and CMa age-metallicity relation (AMR).}
\end{center}
\end{figure*}

\begin{figure*}
\begin{center}
\includegraphics[scale=0.5,angle=-90]{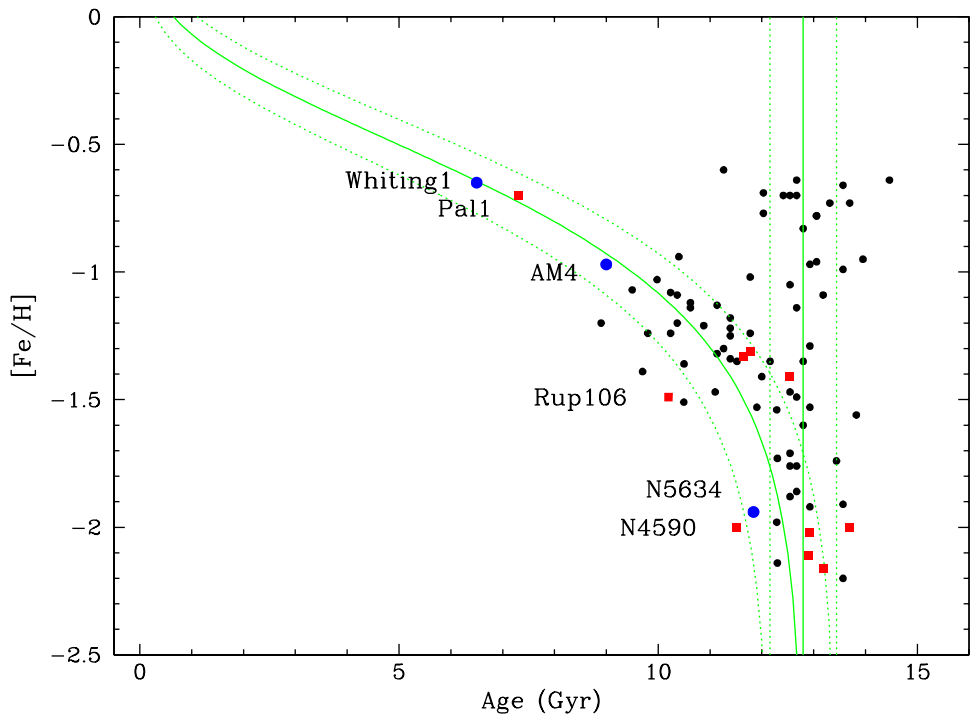}
\caption{Possible globular clusters associated with the Sgr and
  CMa dwarf galaxies. The possible CMa  
globular clusters (red squares) all 
have a similar location in phase space to the 
predicted orbit 
of the CMa dwarf as determined by Martin et al. (2004). 
Three possible Sgr GCs (blue circles) are shown. The 6 additional
GCs that we tentatively include in the Sgr and CMa GC systems are labelled.
Green lines show the joint AMR and the AMR of the metal-poor GCs
in the Milky Way from Marin-Franch et al. (2009). 
}
\end{center}
\end{figure*}

\begin{figure*}
\begin{center}
\includegraphics[scale=0.5,angle=-90]{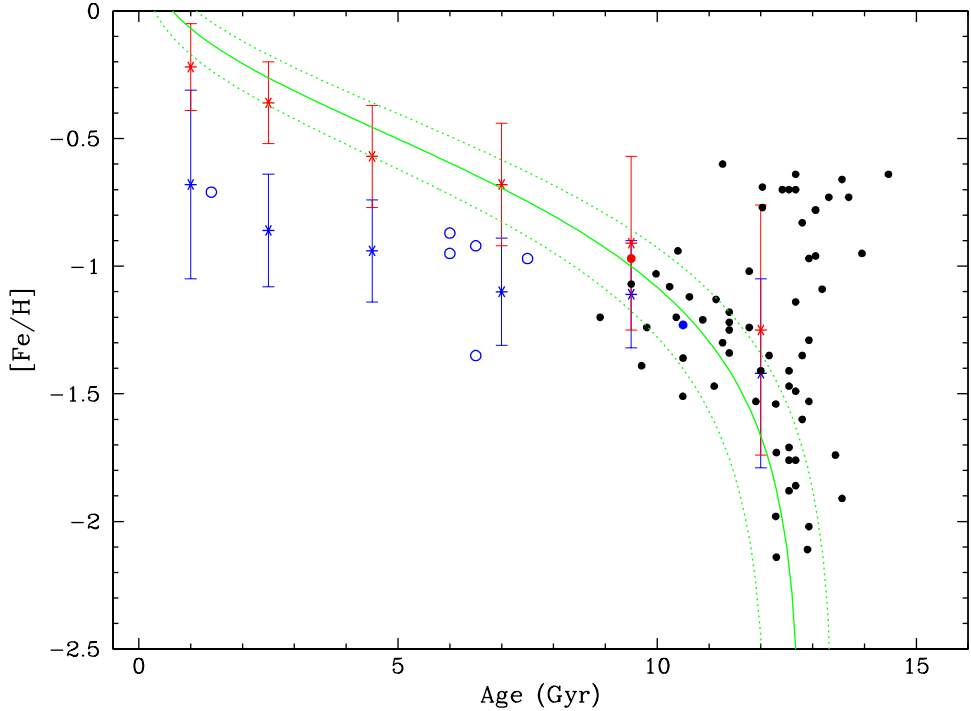}
\caption{Age-metallicity relation of the Magellanic Clouds. The
  mean age and metallicity for the SMC (blue) and LMC disk (red)
  from Carrera et al. (2008a,b) are shown with error bars. SMC
  open and globular clusters from the ACS study of Glatt et
  al. (2008) are shown by open and filled blue circles. The old LMC GC (NGC
121) is shown by a filled red circle (age = 9.5 Gyr, [Fe/H] = --0.97). 
Milky Way globular clusters are shown by filled black circles.
The Sgr and CMa
joint AMR is shown but their GCs have 
been excluded from the plot.}
\end{center}
\end{figure*}

\begin{figure*}
\begin{center}
\includegraphics[scale=0.5,angle=-90]{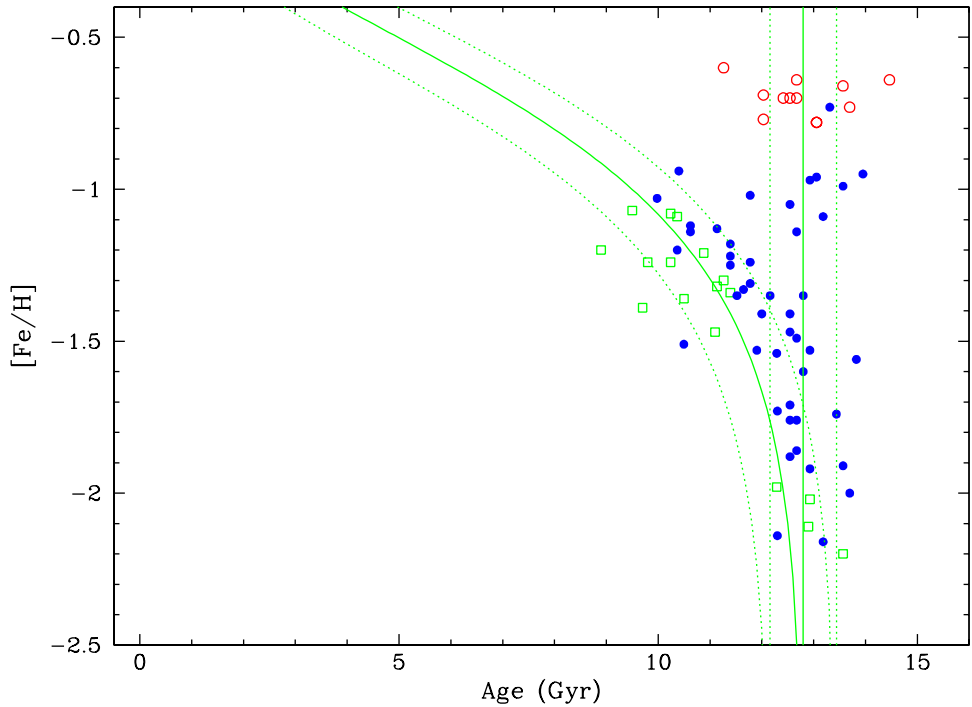}
\caption{Globular clusters coded by horizontal branch (HB) class. 
Bulge/disk are red open circles, old halo are blue filled circles, 
young halo are green open squares. The y axis range has been
reduced compared to previous plots. The Sgr and CMa
joint AMR is shown but their GCs have 
been excluded from the plot. The AMR for metal-poor Milky Way GCs
from Marin-Franch et al. (2009) is also shown.}
\end{center}
\end{figure*}

\begin{figure*}
\begin{center}
\includegraphics[scale=0.5,angle=-90]{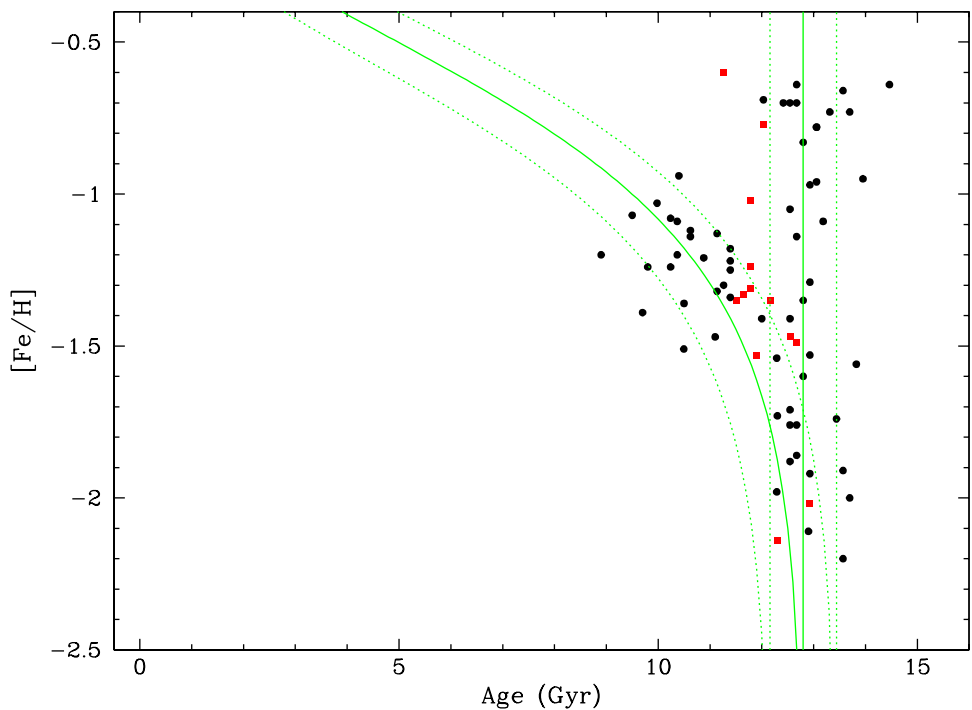}
\caption{Globular clusters with Extended Horizontal Branch (EHB) stars.
GCs with EHBs are shown as red squares. 
Green lines show the Sgr and CMa joint AMR (with their GCs 
excluded) and the metal-poor Milky Way GC AMR.
}
\end{center}
\end{figure*}

\begin{figure*}
\begin{center}
\includegraphics[scale=0.5,angle=-90]{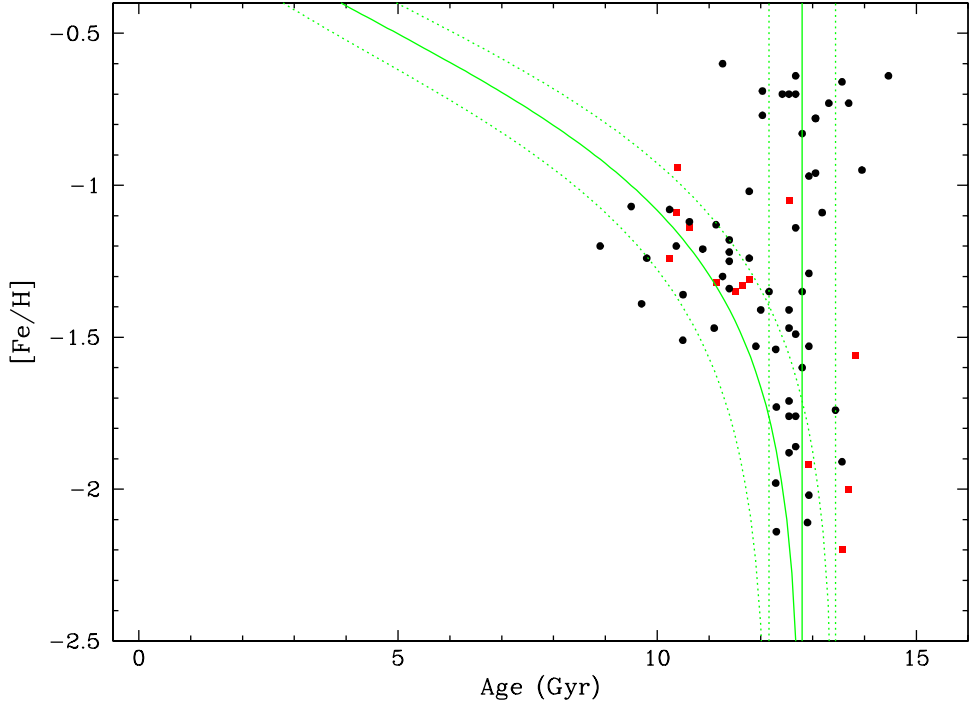}
\caption{Globular clusters known to be 
on retrograde orbits. For many GCs shown the orbital properties are 
unknown. GCs with retrograde motions 
are shown as red squares. 
Green lines show the Sgr and CMa joint AMR (with their GCs 
excluded) and the metal-poor Milky Way GC AMR.
Several of the retrograde motion GCs
are found in the track to younger ages.}
\end{center}
\end{figure*}

\begin{figure*}
\begin{center}
\includegraphics[scale=0.5,angle=-90]{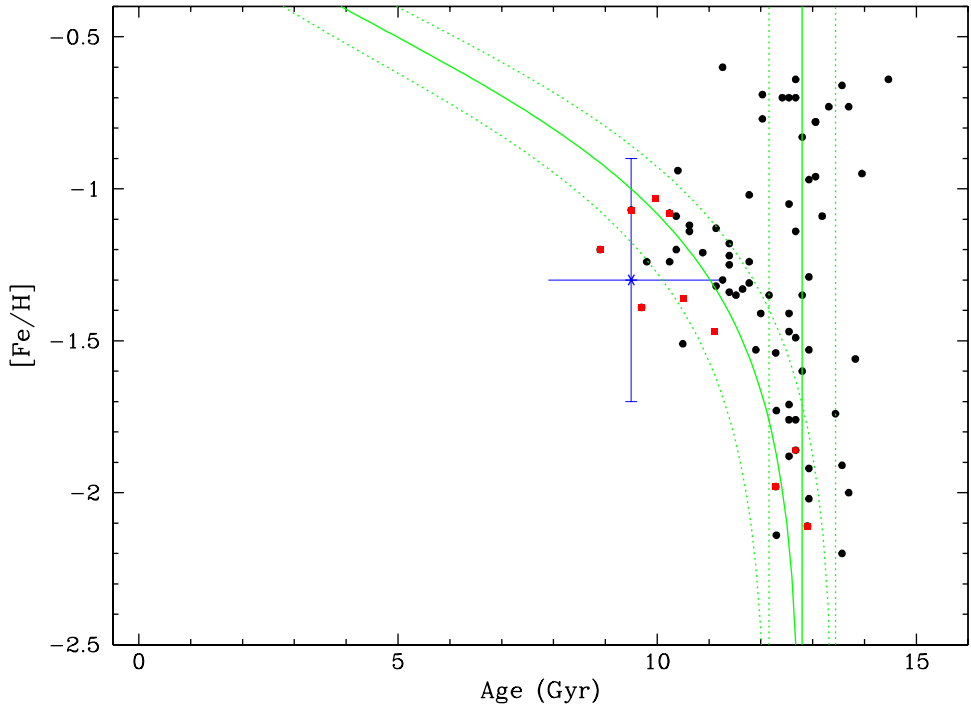}
\caption{Globular clusters coincident with the
Fornax-Leo-Sculptor great circle. The metallicity range of field stars 
formed in the Fornax galaxy $\sim$10 Gyr ago are shown by the
blue cross. GCs that lie the FLS great circle are shown by red squares.  
Milky Way globular clusters are shown by filled black circles.
Green lines show the Sgr and CMa joint AMR (with their GCs 
excluded) and the metal-poor Milky Way GC AMR.
}
\end{center}
\end{figure*}


\end{document}